\documentclass[prb,nofootinbib,twocolumn,superscriptaddress]{revtex4} 

\usepackage{graphicx}
\usepackage{dcolumn}
\usepackage{bm}
\usepackage{threeparttable}
\usepackage{times}
\usepackage{mathptmx}
\usepackage{lscape}
\usepackage{natbib}
\usepackage{amsmath}
\usepackage{amssymb}
\usepackage{braket}

\def\degree{\kern-.2em\r{}\kern-.3em}

\begin{document}


\title{ Compositional asymmetry of disordered structure: Role of spatial constraint }

\author{Koretaka Yuge}
\affiliation{
Department of Materials Science and Engineering,  Kyoto University, Sakyo, Kyoto 606-8501, Japan\\
}%

\begin{abstract}
{  When spatial constraint for the constituents (e.g., atom or particle) of system is once given, disordered structure for \textit{non-interacting} system in equilibrium states is symmetric with respect to equiatomic composition. Meanwhile, when the interaction between constituents is introduced, this symmetry is typically broken, naturally appearing compositional asymmetry. Although this asymmetry, depending on temperature, comes from multibody interactions in the system, we here clarify that the asymmetry near equiatomic composition can be universally well-characterized by two specially selected microscopic structure, which can be known \textit{a priori} without any information about interactions or temperature: The key role is the class of spatial constraint. Based on the facts, we provide analytical expression of temperature dependence of disordered structure, and demonstrate its validity and applicability by predicting short-range order parameters of practical alloys compared with full thermodynamic simulation.
 }
\end{abstract}


\maketitle

\section{Introduction}
In classical systems in equilibrium states, when temperature increases, the system can undergo from ordered to disordered structures due to the competition between interactions between constituents and (configurational) entropy. 
When we prepare set of complete orthonormal basis $\left\{q_1, \ldots, q_g \right\}$ to describe microscopic structures on configuration space under fixed composition $x$, their expectation value at temperature $T$ can be typically obtained through canonical average, $Q_r\left(x,T\right) = Z^{-1}\sum_d q_r^{\left(d\right)} \exp\left(-\beta E^{\left(d\right)}\right)$, where summation is taken over all microscopic states at composition $x$ on phase space. 
Since number of possible microscopic states astronomically increases, potential energy surface should be described by corresponding high-dimentional configuration space. 
Therefore, a variety of calculation techniques have been developed to overcome the practical difficulty, including Monte Carlo (MC) simulation with Metropolis algorism, multicanonical ensembles and entropic sampling to effectively sampling possible microscopic states for predicting macroscopic properties. 
Although the developed approaches have successfully provided accurate prediction of disordered structures in equilibrium states, the role of spatial constraints on composition and temperature dependence of the disordered structures does not get sufficient attentin so far. 

Very recently, we we develop a theoretical approach, enabling to provide new insight into how equilibrium properties (including structures and free energy) connects with spatial constraint on the system.\cite{lsi,emrs,yuge-RM,gps} Through this approach, we find a few special microscopic states, called "Grand Projection states" (GP states), that can be constructed without any information about energy or temperature, can characterize the macroscopic properties. 
In the present study, we extend the approach to investigating composition dependence of disordered structures in equilibrium states near equiatomic composition (i.e., which we call "compositional asymmetry"). 
Through the extention, we provide analytical representation for temperature and composition dependence of disordered structure in terms of the condition of spatial constraint, which can be 
determined by energy of GP states. 

\section{Derivation and discussions}
Let us here consider a binary system (for simplicity), but our derivation can be straightforwardly extended to multicomponent system as seen below.
In classical system under potential energy as a function of spatial positions of constituents, we have found that grand-canonical average of structure can be 
universally given by\cite{gps}
	\begin{eqnarray}
	\label{eq:gps}
	Q_r\left(T\right) \simeq \Braket{Q_r}_{\textrm{1G}} \mp \sqrt{\frac{\pi}{2}} \Braket{Q_r}_{\textrm{2G}} \frac{ I_{\textrm{GP}}^{\left(Q_r\pm\right)} - \Braket{I}_{\textrm{1G}}  }{k_{\textrm{B}}T},
	\end{eqnarray}
where $\Braket{\cdot}_{\textrm{1G}}$ and $\Braket{\cdot}_{\textrm{2G}}$ respectively denotes taking average and standard deviation over all microscopic states on configuration space (including compositions) without weight of Boltzmann factor $\exp\left(-\beta E\right)$. $I$ and $I_{\textrm{GP}}^{\left(Q_r\pm\right)}$ are respectively defined as
	\begin{eqnarray}
	I = E - \Delta \mu N x,
	\end{eqnarray}
and 
	\begin{eqnarray}
	I_{\textrm{GP}}^{\left(Q_r\pm\right)} &=& E_{\textrm{GP}}^{\left(Q_r\pm\right)} - \Delta \mu N \Braket{x}_{\textrm{GP}}^{\left(Q_r\pm\right)} \nonumber \\
	&=& \sum_{t=1}^{s}\Braket{E|Q_t}\Braket{Q_t}_{\textrm{GP}}^{\left(Q_r\pm\right)} - \Delta \mu N \Braket{x}_{\textrm{GP}}^{\left(Q_r\pm\right)},
	\end{eqnarray}
where summation is taken over possible configurational degree of freedom including composition, $\Braket{\cdot}_{\textrm{GP}}^{\left(Q_r\pm\right)}$ denotes partial average over microscopic states satisfying $Q_r \ge \Braket{Q_r}_{\textrm{1G}}$ ($Q_r \le \Braket{Q_r}_{\textrm{1G}}$) for $Q_r+$  ($Q_r-$), and $\Braket{\cdot|\cdot}$ represents inner product on configuration space.
Here, we call $E_{\textrm{GP}}^{\left(Q_r\pm\right)}$ as grand projection (GP) energy along $Q_r$, and corresponding special microscopic structure given by 
$\left\{\Braket{Q_1}_{\textrm{GP}}^{\left(Q_r\pm\right)}, \ldots, \Braket{Q_s}_{\textrm{GP}}^{\left(Q_r\pm\right)}   \right\}$ is called as GP state along $Q_r$, which is 
clearly independent of temperature and energy, and depends only on the class of spatial constraint since $\Braket{Q_t}_{\textrm{GP}}^{\left(Q_r\pm\right)}$ can be obtained by density of microscopic states on configuration space for \textit{non-interacting} system. 
When we choose coordination $Q_r$ as composition, corresponding energy and microscopic states are simply called as GP energy and GP state, which can provide relationship between chemical potential $\Delta \mu$ and grand-canonical average of composition, $x\left(T,\Delta\mu\right)$. 

In the present study, we only focus on the GP energy and state for composition, and we do not derive explicit expression for grand canonical average of structures: The reason is treating 
numerator in Eq.~(\ref{eq:gps}). When we directly apply the previous expression for two dimensional configuration space of $g\left(x, Q_r\right)$ in analogy to our previous approach using characteristics of multidimensional gaussian (here, $g$ denotes density of microscopic states , and $Q_r$ corresponds to pair correlations), 
we can obtain for numerator as
	\begin{eqnarray}
	E_{\textrm{GP}}^{\left(Q_r\pm\right)} - \Braket{E}_{\textrm{1G}} +  \Delta \mu N \cdot \textrm{cov}\left(x, Q_r\right),
	\end{eqnarray}
where $\textrm{cov}\left(x, Q_r\right)$ denotes covariance for $g\left(x, Q_r\right)$. 
Since for even-order correlation should be symmetric at equiatomic composition, we obtain $\textrm{cov}\left(x, Q_r\right) = 0$, which should be only allowed at high temperature limit $T\to \infty$. 
Therefore, in order to apply Eq.~(\ref{eq:gps}) at non-infinite temperature, we should take other stragegies. 
To include the asymmetry of $g\left(x, Q_r\right)$, in analogy to obtaining GP states, we should explicitly take composition-dependent partial average of $x$ in $g\left(x, Q_r\right)$ space. 
Let $\Braket{x}_{\textrm{GP}}^{\left(Q_r\pm\right)}$ be a function of composition $x$. From the constraint condition that  $\Braket{x}_{\textrm{GP}}^{\left(Q_r\pm\right)} = 0$ at $x=\Braket{x}_{\textrm{1G}}$ and $Q_r\left(x,T\right)$ is a quadratic function of $x$ at $T\to\infty$,\cite{sqs} we can determine the composition-dependence, namely, $\Braket{x}_{\textrm{GP}}^{\left(Q_r\pm\right)} = x \Braket{x}^{\left(Q\pm\right)}_{\left(x+\right)} + \left(1-x\right) \Braket{x}^{\left(Q\pm\right)}_{\left(x-\right)} - \Braket{x}_{\textrm{1G}}$.
Here, $\Braket{\cdot}^{\left(Q_r\pm\right)}_{x+}$ ($\Braket{\cdot}^{\left(Q_r\pm\right)}_{x-}$) denotes partial average $\Braket{\cdot}^{\left(Q_r\pm\right)}$ for $x \ge \Braket{x}_{\textrm{1G}}$ 
($x \le \Braket{x}_{\textrm{1G}}$). 
In order to obtain composition dependence of disordered structure from Eq.~(\ref{eq:gps}), we should further determine the relationship between chemical potential $\Delta \mu$ and 
grand-canonical average of composition $x$. This can be easily performed by choosing coordination of $Q_r$ in Eq.~(\ref{eq:gps}) as composition, $x$.\cite{gps}

\begin{figure}[h]
\begin{center}
\includegraphics[width=0.77\linewidth]
{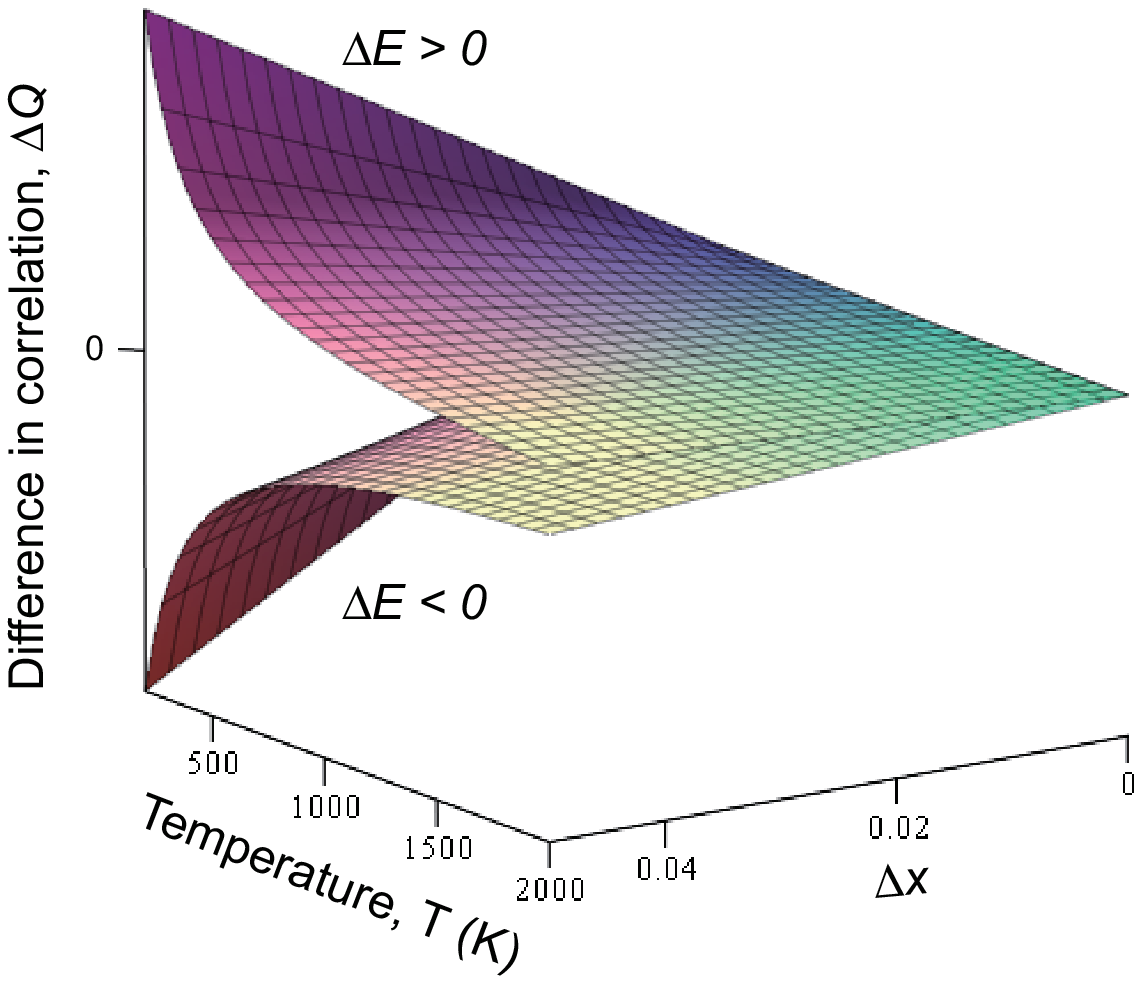}
\caption{Schematic illustration for temperature and composition dependence of difference in structure, $Q_r\left(x + \Braket{x}_{\textrm{1G}},T\right) - Q_r\left(x - \Braket{x}_{\textrm{1G}},T\right)$. $\Delta E$ denotes absolute difference in grand projection energy along composition for $\left(x+\right)$ and $\left(x-\right)$. } 
\label{fig:asym}
\end{center}
\end{figure}
Using the above results, we can give analytical expression for pair correlation near equiatomic composition: 
	\begin{widetext}
	\begin{eqnarray}
	\label{eq:qgc}
	Q_r\left(x,T\right) &\simeq& \Braket{Q_r}_{\textrm{1G}} \mp \sqrt{\frac{\pi}{2}}\Braket{Q_r}_{\textrm{2G}}   \frac{E_{\textrm{GP}}^{\left(Q\pm\right)} - \Braket{E}_{\textrm{1G}}      	\pm    \zeta^{\left(x+/x-\right)}\left(T\right) C^{\left(Q\pm\right)}\left(x\right)   }{k_{\textrm{B}}T} ,
	\end{eqnarray}
where
	\begin{eqnarray}
	\zeta^{\left(x+/x-\right)}\left(T\right) &=& \pm\frac{1}{\Braket{x}^{\left(x+/x-\right)} - \Braket{x}_{\textrm{1G}}} \left\{ k_{\textrm{B}}T \sqrt{\frac{2}{\pi}} \frac{1}{\Braket{x}_{\textrm{2G}}} \left(x - \Braket{x}_{\textrm{1G}} 	\right)  \pm E_{\textrm{GP}}^{\left(x+/x-\right)}  \mp \Braket{E}_{\textrm{1G}}   \right\} \nonumber \\
	C^{\left(Q\pm\right)}\left(x\right)  &=& x \Braket{x}^{\left(Q\pm\right)}_{\left(x+\right)} + \left(1-x\right) \Braket{x}^{\left(Q\pm\right)}_{\left(x-\right)} - \Braket{x}_{\textrm{1G}}.
	\end{eqnarray} 
	\end{widetext}
Here, superscript $\left(x+/x-\right)$ denotes disorderd structure for higher $x$ phase ($x+$) or for lower $x$ phase ($x-$), where their coexistence can be determined from GP energy along $x$, i.e., $E_{\textrm{GP}}^{\left(x+/x-\right)}$.\cite{gps}

From the above equations, we can clearly see that composition dependence of disordered structures can be characterized by five special microscopic states 
(whose energy corresponds to $E_{\textrm{GP}}^{\left(Q_r\pm\right)}$, $E_{\textrm{GP}}^{\left(x+/x-\right)}$ and $\Braket{E}_{\textrm{1G}}$), whose structure can be known 
\textit{a priori}  when spatial constraint on the constituents is given. From Eq.~(\ref{eq:qgc}), we can also see that compositional asymmetry around equiatomic composition is dominated by 
the asymmetry of GP energy along composition, where such asymmetry can be reasonably vanished at high temperature limit of $T\to\infty$. 
This can be schematically shown in Fig.~\ref{fig:asym}, which shows the temperature and composition dependence of difference in structure, $Q_r\left(x + \Braket{x}_{\textrm{1G}},T\right) - Q_r\left(x - \Braket{x}_{\textrm{1G}},T\right)$, for the case of $\Delta E > 0$ and $\Delta E < 0$, where $\Delta E$ denotes absolute difference in grand projection energy along composition for $\left(x+\right)$ and $\left(x-\right)$  (energy is measured from $\Braket{E}_{\textrm{1G}}$).

\begin{figure}[h]
\begin{center}
\includegraphics[width=0.83\linewidth]
{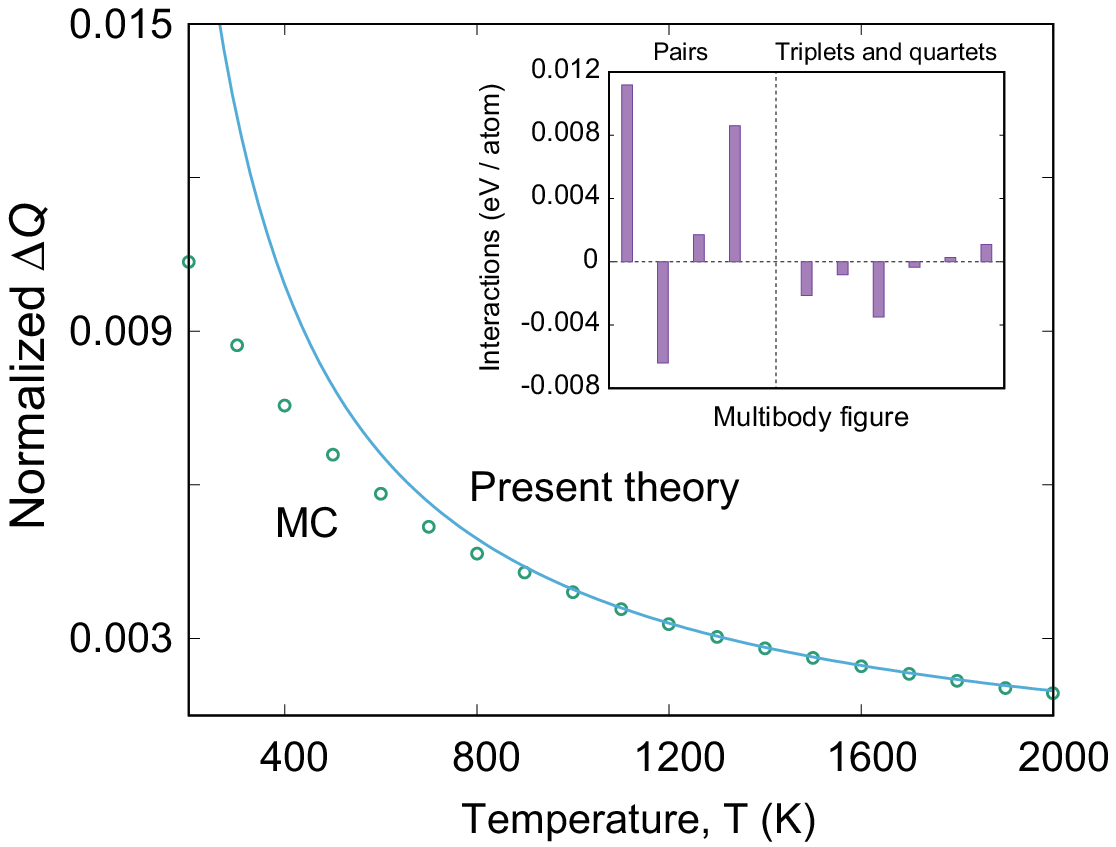}
\caption{Difference in SRO parameter $\Delta Q$ as a function of temperature, predicted by the present theory and MC simulation. $\Delta Q$ is normalized by value of $Q$ that can take maximum under the given spatial constraint, i.e., fcc lattice in this case. Multibody interactions used for MC simulation is given together. }
\label{fig:dq}
\end{center}
\end{figure}

In order to confirm the tendency of compositional asymmetry, we compare the results given in Fig.~\ref{fig:asym} with those obtained by full thermodynamic simulation: We artificially prepare effective multibody interactions in terms of generalized Ising model on fcc lattice with binary elements, which provides well-known ordered structure of "40" at the ground state with order-disordere transition temperature is around 150 K.\cite{emrs} We estimate difference in short-range order (SRO) parameter $\Delta Q$ for the system between $x=0.46$ and $x=0.54$, which is symmetric with respect to equiatomic composition of $x=0.5$. 
Temperture dependence of SRO is quantitatively estimated by applying the multibody interactions to Monte Carlo (MC) statistical simulation under canonical ensemble, where the MC cell contains 2048 atoms (i.e., $8\times 8\times 8$ expantion of conventional fcc unit cell) with 8000 MC step per site to take ensemble average.  The predicted $\Delta Q$ by the present theory and MC simulation is shown in 
Fig.~\ref{fig:dq} together with the multibody interaction used. We can clearly see that at high temperature above $\sim800$ K, SRO by the present theory exhibit excellent agreement with that by MC, while it shows deviation with decrease of temperature. This deviation can be reasonablly interpreted since our theory is based on the configurational density of states (CDOS) for \textit{non-interacting} system well-characterized by multidimensional gaussian, whose deviation from practical CDOS should be naturally enhanced by Boltzmann factor $\exp\left(-\beta E\right)$ at low temperature with the foot of the CDOS from its center of gravity where effect of spatial constraint on CDOS, especially information about landscape of higher order moment (typically, greater than two), plays significant role, which has already been confirmed by our previous studies. Inclusion of information about higher-order moments of CDOS into the proposed Eq.~(\ref{eq:qgc}) therefore should be our future study.


\section{Conclusions}
By focusing on the role of spatial constraint on equilibrium properties, we propose analytical representation for compositional asymmetry of disordered structure in binary system, which is dominated by energy of two specially selected microscopic states whose structures can be known a priori without any information about energy or temperature. We demonstrate the validity of the proposed representation by predicting the short-range order tendency on fcc lattice, compared with full thermodynamic simulation based on generalized Ising model: While deviation is enhanced with decrease of temperature, we find excellent agreement at high temperature region.

\section*{Acknowledgement}
This work was supported by a Grant-in-Aid for Scientific Research (16K06704) from the MEXT of Japan, Research Grant from Hitachi Metals$\cdot$Materials Science Foundation, and Advanced Low Carbon Technology Research and Development Program of the Japan Science and Technology Agency (JST).

\end{document}